\documentclass[12pt,preprint]{aastex}
\usepackage{epsfig}

\begin{document}

\title{Discovery of Two Types of X-ray Dips in Cyg X-1}
\author{Y.X. Feng and Wei Cui}
\affil{Department of Physics, Purdue University, West Lafayette, 
IN 47907}
\email{fengyx@physics.purdue.edu}
\email{cui@physics.purdue.edu}

\begin{abstract}

We observed Cyg X-1 with {\em RXTE} contiguously over its 5.6-day binary 
orbit. The source was found to be in the hard state throughout the 
observation. Many intensity dips were detected in the X-ray light curves. 
We found that the dips fell into two distinct categories based on their 
spectral properties. One type exhibits strong energy-dependent attenuation 
of X-ray emission at low energies during a dip, which is characteristic 
of photoelectric absorption, but the other type shows nearly 
energy-independent attenuation. While the first type of dips are likely 
caused by density enhancement in an inhomogeneous wind of the companion 
star, as previous studies have indicated, the second type might be due 
to partial obscuration of an extended X-ray emitting region by optically 
thick ``clumps'' in the accretion flow. It is also possible that the 
latter are caused by a momentary decrease in the X-ray luminosity of the 
source, due, for instance, to a decrease in the mass accretion rate, or
by Thomson scattering in highly ionized ``clumps''. We discuss the 
implications of these scenarios.

\end{abstract}


\section{Introduction}

Cyg X-1 was the first astronomical system discovered to show strong 
evidence for
a stellar mass black hole (Bolton 1972; Webster $\&$ Murdin 1972). 
It is a binary system with an orbital period of 5.6 days. The results 
from radial velocity measurements imply that the mass of the compact 
object exceeds the upper limit on the mass of a neutron star, so the 
compact object is inferred to be a black hole. A more recent study, based 
on spectrum synthesis, has derived a mass of about 10 M$_{\odot}$ for the 
black hole (Herrero et al. 1995). The companion star has been identified 
as an O9.7 Iab supergiant (Walborn 1973; Gies $\&$ Bolton 1986), with a 
mass of about 20 M$_{\odot}$ (Herrero et al. 1995). Cyg X-1 is, therefore, 
intrinsically different from most known black hole candidates which 
contain only a low-mass companion star. This difference might be related 
to the fact that Cyg X-1 is a persistent X-ray source while those with a 
low-mass companion are exclusively transient sources. 

The X-ray emission from Cyg X-1 is likely powered by accretion of material
from the companion star by the black hole. In this case, the accretion 
flows are thought to follow a pattern that is intermediate between that 
of Roche-lobe overflow, as in low-mass systems, and that of wind 
accretion, which is common for high-mass systems (Gies $\&$ Bolton 1986). 
Such an accretion process is sometimes referred to as ``focused wind 
accretion'', which can occur when the companion star is close to filling 
its Roche lobe (Friend \& Castor 1982). The X-ray observations of Cyg X-1 
have revealed that it has two distinct spectral states, hard and 
soft (see reviews by Oda 1977, Liang $\&$ Nolan 1984, Tanaka \& Lewin 
1995, Cui 1998, and Liang 1998).  The source is usually found in the hard 
state, when its X-ray spectrum is relatively flat (with a typical power-law 
photon index of 1.5), but, occasionally, it makes a transition to the soft 
state, when the spectrum steepen significantly (to a typical photon index 
of 2.5). 

In the hard state, the X-ray intensity of Cyg X-1 is strongly modulated by 
the binary motion (Wen et al. 1999; Brocksopp et al. 1999; Priedhorsky, 
Brandt, $\&$ Lund 1995; Holt et al. 1979), probably caused by varying amount 
of absorbing material along the line of sight through the stellar wind of 
the companion star. As expected, the intensity is found to reach a minimum 
at the times of superior conjunction of the black hole (when the companion 
star is in front of the black hole with respect to the line of sight), which 
is usually defined as orbital phase $0$. It is, however, a very broad 
minimum, spanning about 26$\%$ of the orbit. In contrast, no X-ray orbital 
modulation is measurable in the soft state (Wen et al. 1999), which might be 
due to a significant change in the physical conditions of the wind (e.g., 
much higher ionization level) and/or in the geometry of the X-ray emitting 
region.

Besides the global orbital modulation of the X-ray emission, X-ray intensity 
dips are often seen in the light curves of Cyg X-1 (e.g., Pravdo et al. 1980; 
Remillard \& Canizares 1984; Kitamoto et al. 1984; Ba{\l}uci{\'n}ska \& 
Hasinger 
1991; Ebisawa et al. 1996). The dips vary in duration from minutes to hours. 
The spectrum of the source hardens during a dip, implying that the dips 
probably originate in the photoelectric absorption of X-rays by the ambient 
medium. This is further supported by the detection of Fe K absorption edge in 
the dip spectrum in some cases (Kitamoto et al. 1984). The absorbing column 
density can increase by more than one order of magnitude during a dip 
(Kitamoto et al. 1984; Ba{\l}uci{\'n}ska-Church et al. 1997). The distribution 
of dips against orbital phases shows a prominent peak around phase $0$ 
(Ba{\l}uci{\'n}ska-Church et al. 2000), implying an origin of the dips in the 
stellar wind from the companion star.

In this paper, we present results from a long observation of Cyg X-1 
with the {\em Rossi X-Ray Timing Explorer} (RXTE) over one complete 
5.6-day orbital cycle. Many intensity dips were detected. The large 
collecting area of the Proportional Counter Array (PCA) aboard {\em RXTE} 
made it possible to conduct a more detailed investigation of spectral 
evolution of the source during some of the strong dips, as well as to 
carry out orbital-phase-resolved spectroscopy to quantify the variation 
in the column density over an orbital cycle.

\section{Data}

For six consecutive days (2000 January 5--11) Cyg X-1 was observed with
the large-area detectors abroad RXTE, with usual interruptions because of earth 
occultation and passage of the satellite through the South Atlantic 
Anomaly. The data were collected for an effective exposure time of about 
240 ks. Judging from the long-term ASM/RXTE light curves of the source
(both the flux and hardness ratios), Cyg X-1 was in its usual hard state 
when our pointed RXTE observations were made.

Besides the PCA, {\em RXTE} also
carries another pointing instrument, the High-Energy X-ray Timing 
Experiment, which covers a nominal energy range of 15--250 keV. For 
this investigation, only the PCA data (2--60 keV) were used, because 
we are primarily interested in spectral changes at low energies. The 
PCA consists of five {\em Proportional Counter Units} (PCUs). 
Not all PCUs were turned on throughout the observation, because of technical 
difficulties with some of the PCUs. Consequently, the number of PCUs 
used varied between 3 and 5. Because of this, it was necessary to 
break up the observation into a number of segments and to analyze
data for each segment separately.

\section{Analyses and Results}

Both spectral and timing analyses were carried out using the 
{\it Standard 2} data (which has a timing resolution of 16 seconds).
We used {\em ftools v5.0}, as well as the calibration files 
and background models that accompanied this release of the software, 
to extract X-ray spectra and light curves of Cyg X-1. The quality of 
the background models was examined by comparing an overall spectrum 
(source plus background) with the corresponding model background 
spectrum. The two should overlap each other at the highest 
energy channels, since the counts in those channels should be entirely 
due to the background. We found that background modeling was 
satisfactory, except for a couple of cases where renormalization of 
the background was necessary.

\subsection{X-ray Dips}

We made a light curve for each of the following energy bands: 3--6 keV, 
6--15 keV, and 15--60 keV, using data from all PCUs (and all xenon 
layers) that were turned on for a segment of the observation. A 
corresponding background light curve was then constructed from the 
models and was subtracted from the overall light curve to obtain the 
light curve of the source for a particular energy band. The light curve 
of each segment was then normalized to count rates per PCU, and all 
segments were pieced together to obtain the final light curve for the 
entire observation. We also computed hardness ratios by taking the 
ratio of the light curves in two energy bands (higher energy band over 
lower energy band). The hardness ratios are useful in providing crude 
information about the spectral properties of a source.

The light curves of Cyg X-1 reveal frequent occurrences of a brief 
decrease in the intensity of the source. These are X-ray dips. In 
most cases, the hardness ratios indicate that the spectrum of Cyg X-1 
hardens during a dip, implying that the dips are caused by additional 
absorption. As an example, Figure.~1 shows a segment of the light curves 
where two dips are clearly visible (on top of an X-ray flare). As a 
matter of fact, the 
dips manifest themselves much more prominently as peaks in the hardness 
ratio plots, which are also shown in the figure. Note that we have 
plotted orbital phases instead of times to emphasize the distribution 
of dips over the orbital cycle. For this purpose, we have used the 
most updated ephemeris of Cyg X-1 (Brocksopp et al. 1999). The range 
of orbital phases covered by our observations is between 0.50 and 1.60.

To obtain a complete sample of dips, we used the 3--6 keV light curve 
alone, because it is most sensitive to absorption dips and the data is 
of the highest quality (compared to light curves in higher energy bands).
We adopted the following algorithm to detect dips. First, we smooth
the light curve by a running boxcar with a width of 3 time bins (or
48 s), to minimize intrinsic variability of the source. Second, we 
scan over the entire light curve for all ``local minima''. A local 
minimum is defined as a point with two higher adjacent neighbors. 
Third, we compare each local minimum with its next closest neighbor on 
each side. This process continues until two points on each side are 
found to be more than 20$\sigma$ higher than the local minimum itself 
(where $\sigma$ is the error bar of the local minimum). A dip is then 
identified. On the other hand, the process stops, if one neighbor on 
either side is lower than the local minimum or if the time interval 
examined exceeds 300 time bins (or 4.8 ks) or reaches the overall
length of a data segment, before the 
20$\sigma$ threshold is reached. We then move onto the next local 
minimum, until all minima in the list have been examined. The choice 
of a relatively high detection threshold is to minimize confusion due 
to the intrinsic variability of the source. Since the main objective 
of this investigation is to study the spectral properties of dips in a 
more detailed manner than what has been done previously (rather than
focusing on the frequency of the dips; see
(Ba{\l}uci{\'n}ska-Church et al. 2000), the reliability of detection 
is much more important than the number of dips that we can detect. We 
note that our algorithm is clearly biased against dips of very short 
durations (less than 80 s) or those of very long durations (greater 
than 4.8 ks or the duration of a data segment).
However, it is very difficult to improve the situation, because the
detection of short-duration dips is limited by the intrinsic 
variability of the source and the detection of long-duration ones is 
complicated by coverage gaps in the data.

We detected a total of 33 dips. To examine their distribution over 
the binary orbit, we divided the orbit into 20 phase bins and computed
the {\em normalized} number of dips in each phase bin (which is the 
number of dips detected divided by the fractional coverage of the phase 
bin). The normalization is necessary because there are gaps in the 
light curve and thus the coverage is incomplete for all phase bins
(note that each phase bin corresponds to about 0.28 days or 6.7 hours).
The results are shown in Figure~2. The overall shape of the distribution, 
such as the apparent asymmetry about phase $0$, is in general agreement
with that derived previously (Ba{\l}uci{\'n}ska-Church et al. 2000), 
although the statistics are very limited here. The asymmetry in the 
distribution might be a manifestation of enhanced ``clump'' formation 
in the leading hemisphere of the stellar wind, because of the compression 
of the wind by orbital motion.

The duration of the dips varies greatly. Figure 3 shows 
a dip that has the longest duration among all detected. It should be
noted that this dip was {\em not} detected by the algorithm, because 
of its long duration as well as the data gaps present.
It is interesting that it occurred almost exactly 
at phase $0$. Unfortunately, the presence of gaps in the data makes it 
impossible to derive its duration accurately. The profile of a dip can 
be quite complex, certainly not necessarily symmetric. Figure 4 shows an
example of such a dip. Sub-structures within the dip are quite apparent
(although it can also be argued that it is a superposition of two 
smaller dips).

Upon a closer examination of the dips detected, we found that they
fell into two distinct categories, based on their spectral properties. 
While the majority of the dips show strong spectral variation over the 
duration of dipping activity (see the hardness ratio plots in Figures 1, 3, 
and 4), there are some dips that show little spectral variation. We 
refer to the former as type A and the latter as type B. As examples, 
Figures 5 and 6 show the light curve and hardness ratios of two type B 
dips of quite different durations. It is clear that type B dips can be
comparable in depth to type A ones (e.g., comparing Figure 1 and Figure 6)
but they show much less spectral variation (if any at all). Of the 33 
dips detected, 4 dips are quite convincingly type B. They seem to 
distribute randomly over the orbit: one is around phase 0.637, one 
around phase 0.830, one around phase 0.995,  and one around 1.148, but
we do not have sufficient statistics to address this issue with 
confidence. 

\subsection{Dip Spectra}

We proceeded to carry out more detailed spectral analyses and modeling 
for some of the strong dips. We divided the overall light curve into 
dip and non-dip time intervals (with gaps between the two to ensure the 
quality of the dip spectra). We then extracted a source spectrum for 
each time interval by combining data from all PCUs (and all xenon 
layers) that were on during the observation. 
To strengthen some of the conclusions drawn from the hardness ratios in 
a model-independent manner, we simply divided the spectrum of the source
during a dip by the average non-dip spectrum. For illustration, Figure 7 
shows the results for a type A dip and a type B dip, respectively. 

We can see that the spectrum of the type A dip rolls over much faster 
than the non-dip spectrum below about 6 keV, indicating the need for 
additional absorbing material intrinsic to the binary system. It is 
interesting to notice that the dip spectrum never fully recovers to 
the non-dip spectrum even at high energies. This seems to signify the 
importance of Thompson scattering in attenuating X-ray flux (at a 
few percent level). The presence of sufficient number of electrons to 
participate in the scattering process would imply that the absorbing 
medium is perhaps highly ionized, which has been suggested previously 
(Pravdo et al. 1980; however, see Kitamoto et al. 1984; 
Ba{\l}uci{\'n}ska-Church et al. 1997).

To be more quantitative, we fit the spectra with an empirical model 
that was used in previous studies of Cyg X-1 (Ebisawa et al. 1996; 
Cui et al. 1997). First, we fit the non-dip spectrum with a broken 
power law (plus a Gaussian function, which is needed to minimize 
residuals between 6--7 keV). A detailed discussion of physical models 
is beyond the scope of this paper. Here we only focus our attention 
on how the column density varies in a dip. The model describes the 
spectrum quite well. The inferred column density 
is $0.8 \pm 0.4 \times 10^{22}\mbox{ }$cm$^{-2}$, which is in general
agreement with previous measurements (e.g., Ba{\l}uci{\'n}ska \& Hasinger 
1991; Ebisawa et al. 1986; Cui et al. 1998). We then applied the 
best-fit model for the non-dip spectrum to the dip spectrum by fixing
all model parameters except for the column density. This procedure is 
necessary because the quality of the dip spectrum is not adequate for 
constraining so many parameters in the model, but there is no physical
justification for doing so. This only represents an attempt to 
quantify the amount material in the binary system that caused an
absorption dip. The model fits the dip spectrum quite well. The 
inferred column density for the dip is
$1.7 \pm 0.1 \times 10^{22}\mbox{ }$cm$^{-2}$, which roughly 
doubles the interstellar value. We note that this is only the average 
column density for the dip; the peak value can be much higher. 

The situation for type B dips is quite different. For the type B dip
shown in Figure 5, the ratio of the dip and non-dip spectra is nearly 
flat up to at least 20 keV, above which the statistics become quite 
poor. If we average all data points above 20 keV, the average level 
seems to lie above that for data points below 20 keV. The lack of a 
strong energy dependence of the ratio at low energies indicates that 
the observed attenuation cannot be due to  photoelectric 
absorption. There are at least three possible scenarios that might 
explain the type B dips. The first scenario attributes the type B dips 
to partial covering of an extended X-ray emitting region by material 
that is totally opaque. The second one is that the source experiences 
a momentary decrease in its luminosity during a type B dip, due, for 
instance, to a sudden decrease in the mass accretion rate. The third 
possibility is that the type B dips originate in pure Thompson 
scattering, implying that the clumps must be almost fully ionized. 

\subsection{Orbital Variation in Column Density}

To investigate the global orbital modulation of X-ray emission previously 
seen (e.g., Wen et al. 1999), we binned the observation into 10 orbital 
phases. For each orbital phase, we extracted a spectrum of the source, as 
before, but used data only from the first xenon layer of each PCU to 
minimize calibration uncertainties, since it is the most accurately calibrated 
among all xenon layers. The trade-off is that we lost spectral coverage at 
energies above about 25 keV. We can afford such a loss here, because we 
are only interested in spectral variation at the lowest energies due to 
photoelectric absorption.

For each orbital phase, we fit the spectrum with the same model described 
in the previous section (broken power law plus Gaussian function) and 
the results are satisfactory. Figure 8 shows the inferred column density 
at different orbital phases. Although the column density seems to be 
higher near superior conjunction, large uncertainties prevent us from 
drawing a definitive conclusion, especially in the presence of other high 
points, e.g., at phase 1.3, which may or may not be real. The large 
uncertainty in determining the column density arises mostly from the fact 
that the sensitivity of the PCA drops precipitously below about 3 keV.
Ba{\l}uci{\'n}ska-Church et al. (2000) computed variation in column density 
through the stellar wind due to orbital motion, based on a simple wind 
geometry, for two extreme cases with and without the suppression of 
radiatively-driven wind. Our results seem to lie between those two cases 
(comparing Figure 8 with Figure 7 of Ba{\l}uci{\'n}ska-Church et al. 2000).  

\section{Discussion}

As mentioned in the introduction, the phenomenon of X-ray intensity dips 
in Cyg X-1 is well known and well studied. Over the years, a rich 
database has been built up for this famous black hole candidate. The
database is essential for a systematic study of dips, whose occurrence
is somewhat unpredictable and is thus easy to miss in a brief pointed
observation. The distribution of the dips over a binary orbital cycle 
has been reliably established recently (Ba{\l}uci{\'n}ska-Church et al. 
2000): most of the dips occur near superior conjunction of the X-ray 
source, but dipping activity has also been seen at orbital phases far 
from superior conjunction. Our results are consistent with that. We 
went a step further in this study. We carried out a long observation of
Cyg X-1 over one entire orbital period with the large-area detectors
aboard {\em RXTE}. The data allowed us to conduct more detailed 
spectral analyses of the dipping phenomenon as well as phase-resolved 
spectroscopy to quantify the variation in the column density over an 
orbit.

The main result of this investigation is the recognition of the
existence of two types of dips. The spectrum of type A dips shows 
an energy-dependent reduction at low energies that is characteristic of 
photoelectric absorption. Detailed spectral modeling shows that the only 
difference between the spectrum of type A dips and the average non-dip 
spectrum seems to be the presence of additional column density during a 
dip. We stress, however, that our data is not of sufficient quality to 
warrant more complicated modeling, as has been done previously (e.g., 
Ebisawa et al. 1996). The type A dips seem to occur preferentially
around the times of superior conjunction of the black hole. The 
spectrum of type B dips, on the other hand, shows almost an 
energy-independent reduction at low energies. Therefore, these dips 
cannot possibly be due to  photoelectric absorption. Moreover, 
the type B dips appear to distribute randomly over the binary orbit.

Type A dips are much more common than type B dips. They are probably 
produced by density enhancement in an inhomogeneous wind from the 
companion star, since the column density changes only moderately during 
such a dip. They occur more frequently around superior conjunction 
probably because the line-of-sight follows a longer path through the 
denser part of the wind (however, see Blondin \& Woo 1995). The 
inhomogeneities in the wind can be caused by a 
variety of physical processes (see Ba{\l}uci{\'n}ska-Church et al. 2000 
for a few examples). It remains to be seen whether the observed global 
orbital modulation of X-ray emission is entirely due to the presence 
of such dips. 

Type B dips might be caused by partial covering of an extended X-ray 
emission region by an opaque ``screen''. In the context of 
Comptonization models (e.g., review by Liang 1998 
and references therein), the presence of an extended emitting region 
may not be unreasonable for Cyg X-1. The question is what could 
physically serve as an opaque ``screen'' to produce the observed type B 
dips. The answer may lie in the fact that 
Cyg X-1 is a high-mass X-ray binary (HMXB), in which the mass accretion 
process is probably mediated by stellar wind from the massive companion 
star. Numerical simulations of mass accretion in HMXBs show that a 
tidal stream develops when the companion star is close to filling its 
Roche lobe and that the stream trails behind the compact object due 
to the Coriolis force (Blondin et al. 1991). The density in the stream
is shown to be as high as 20--30 times the ambient wind density. Since
the obscuration of X-rays occurs close to the accretion disk in this
case, it is not expected to be very sensitive to the binary motion. Now, 
the question is whether this scenario can be applied to Cyg X-1, whose 
orbit is only moderately inclined, with a probable inclination angle 
of 30\arcdeg -- 40\arcdeg\ (Bolton 1975; Guinan et al. 1979; Daniel 1981). 
A full three-dimensional simulation would be required to show the scale 
height of the tidal streams. We note that the inclination angle remains 
to be poorly determined for Cyg X-1, which is one of the largest 
contributors to the uncertainty in determining the mass of the compact 
object. For instance, the optical polarization measurements imply a 
wide range of 25\arcdeg -- 70\arcdeg (e.g., Long et al. 1980). 

Type B dips might also be due to a sudden decrease in the mass accretion 
rate, causing a decrease in the X-ray luminosity of the source itself. 
This could, in principle, occur, since, after all, the intrinsic 
variability of the source is generally thought to be due to fluctuation 
in the mass accretion rate. However, for Cyg X-1 (and black hole 
candidates in general), a change in X-ray flux is generally accompanied
by a change in spectral hardness. In fact, it has been shown that in the 
hard state the flux and spectral hardness are, to some degree, 
anti-correlated (e.g., Wen et al. 2001). However, the correlation 
is quite weak, so our results cannot rule out this possibility. 

It is also possible that type B dips are produced by (nearly) pure 
Thomson scattering of X-rays in highly ionized dense clumps in the wind. 
Given that the cross section for photoelectric absorption is, on average, 
more than two orders of magnitude larger than the Thomson cross section, 
the density of a clump that produces a type B dip is required to be more 
than two orders of magnitude larger than that of a clump that produces 
a type A dip of comparable depth (assuming that the clumps are comparable 
in size). To abtain an ionization parameter that is, for
instance, a factor of 100 larger for type B clumps than for type A 
clumps, we would then require that the former are, on average, more than
100 times closer to the X-ray source. If so, this scenario could also 
explain why the type B dips are much less dependent of orbital phases.

Finally, we made an attempt to quantify how the column density varies 
with orbital motion by carrying out orbital-phase-resolved spectroscopy.
The results seem to indicate that the column density is higher at
superior conjunction, although error bars are too large for us to be
definitive about it. The issue can be resolved by a similar observation 
with detectors of much improved low-energy response and spectral 
resolution, such as those
aboard {\em Chandra} and {\em XMM}. Much improved spectral resolution
of these detectors will also allow the detection of absorption edges, 
which provides means of determining column density in a manner that
is independent of continuum modeling (Schulz et al. 2001). Furthermore,
the highly eccentric orbits of these satellites allow long un-interrupted
observations, which is very much important for studying the distribution 
of dips over a binary orbit.

\acknowledgments
We wish to thank Dr. Douglas Gies, the referee, for many helpful comments. 
In particular, he suggested that type B dips might be caused by a change 
in the luminosity of the source and that the asymmetry in the dip 
distribution might be due to the compression of the stellar wind by orbital 
motion. This work was supported in part by NASA through grants NAG5-9098 
and NAG5-9998.

\clearpage
\begin{figure}
\centerline{ 
\psfig{figure=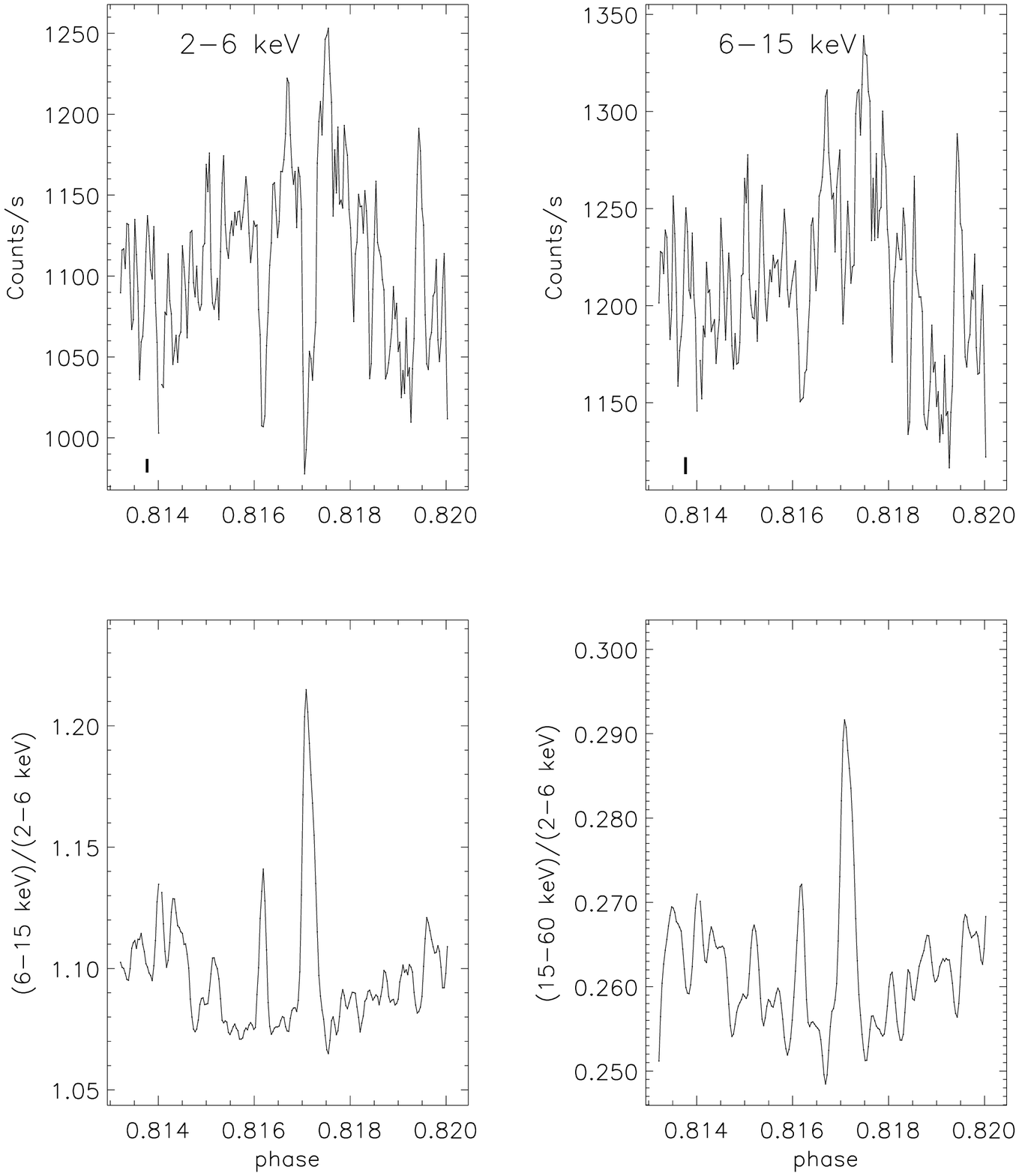,width=1.1\textwidth,angle=0} }
\caption{X-ray dips in Cyg X-1. The time bin of each data point is 16 s. 
The light curves have been smoothed by a running boxcar of a width of 3 
time bins. The typical size of error bars is shown for each light curve. }
\end{figure}
\clearpage
\begin{figure}
\centerline{
\psfig{figure=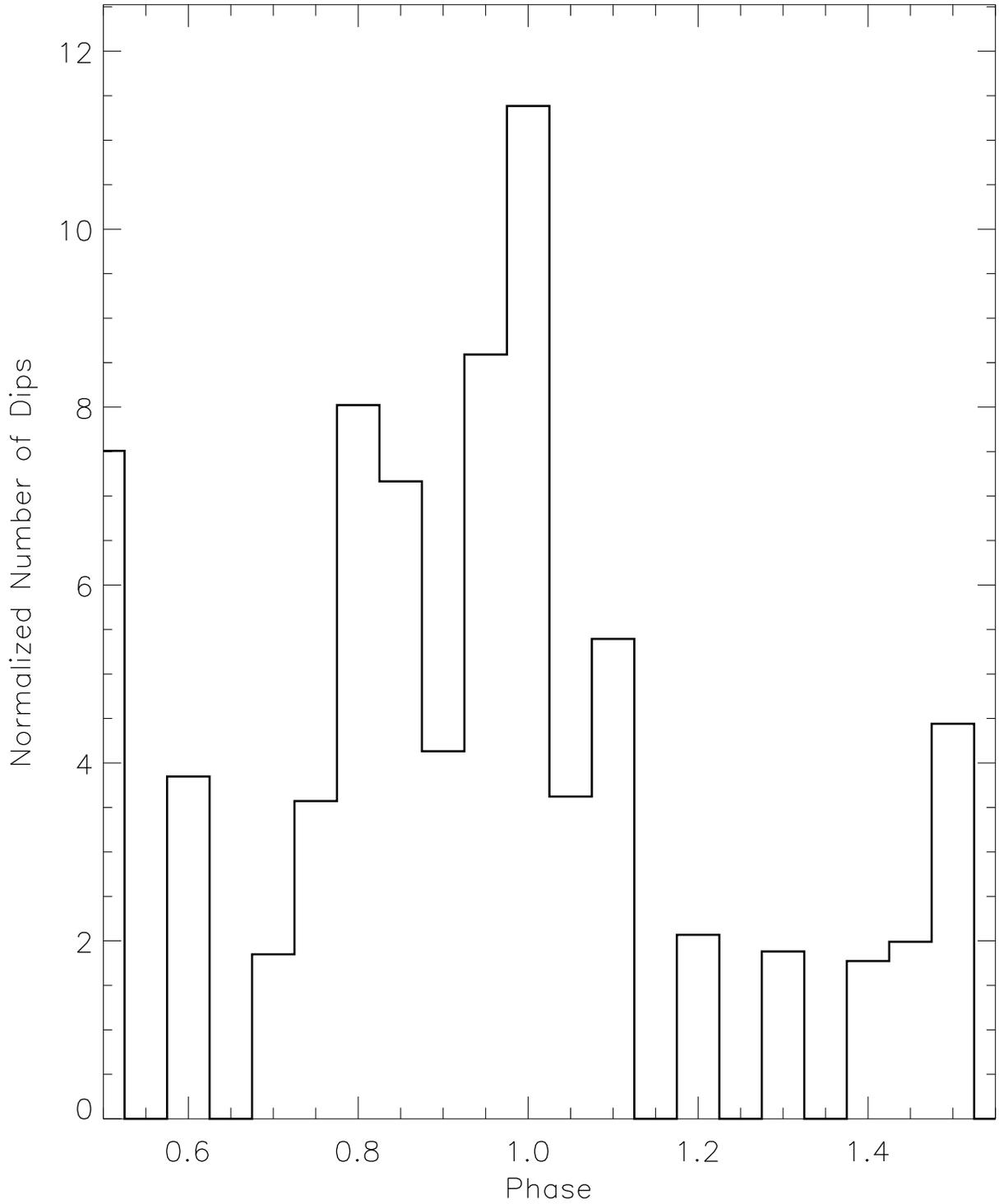,width=1.1\textwidth,angle=0} }
\caption{Orbital distribution of dips in Cyg X-1. The normalized number
of dips in a phase bin is derived by dividing the actual number of dips 
seen in the phase bin by the fractional coverage of the bin (i.e., it 
represents the number of dips that would have been detected had the 
phase bin been fully covered by the observation). }
\end{figure}
 \clearpage
\begin{figure}
 \centerline{
\psfig{figure=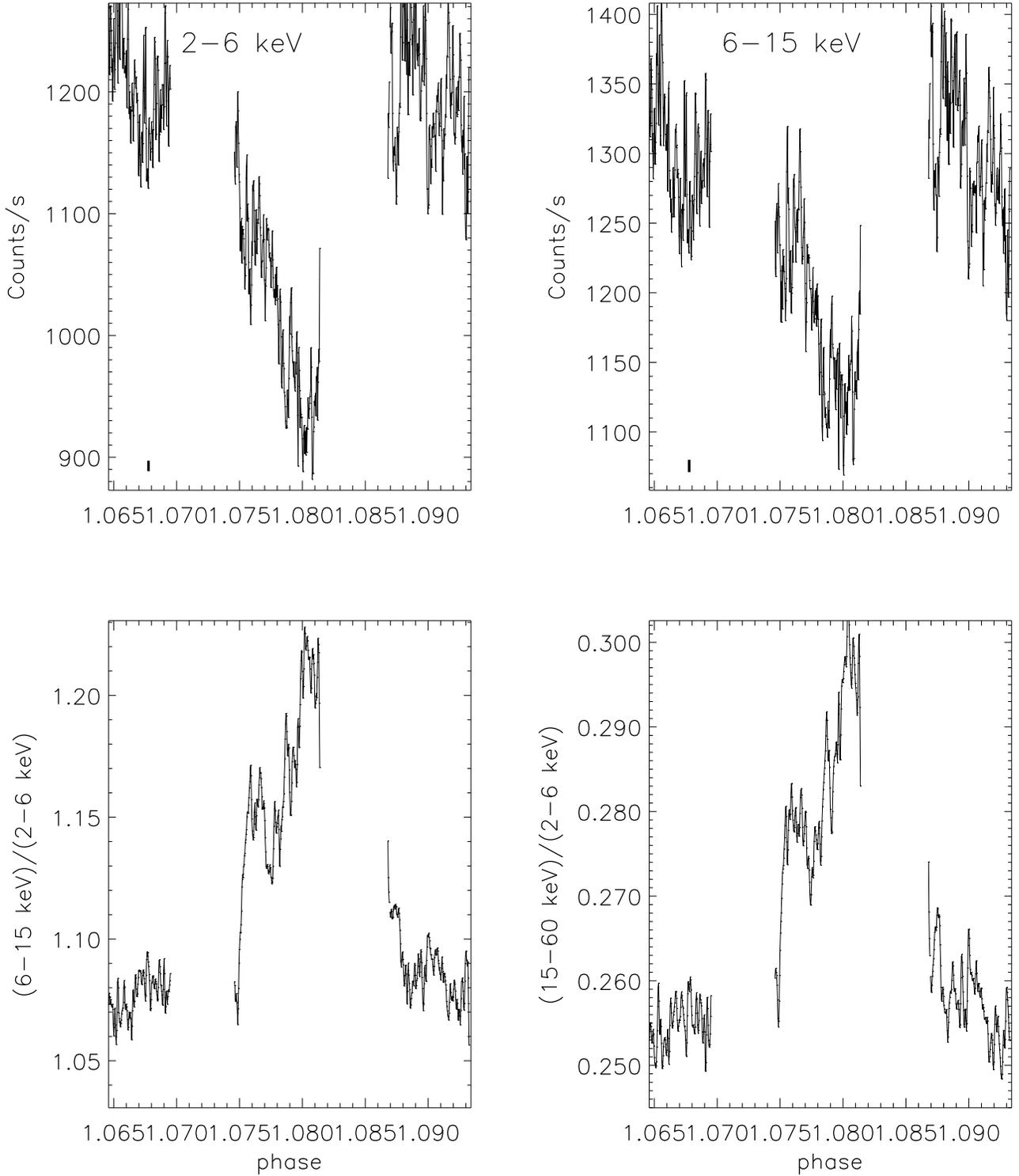,width=1.1\textwidth,angle=0} }
\caption{Same as Figure 1, but for the longest dip detected (by visual
inspection; see text). Note that it occurred almost exactly at the time 
of superior conjunction of the black hole and lasted for more than an 
hour. }
\end{figure}
 \clearpage
\begin{figure}
\centerline{ 
\psfig{figure=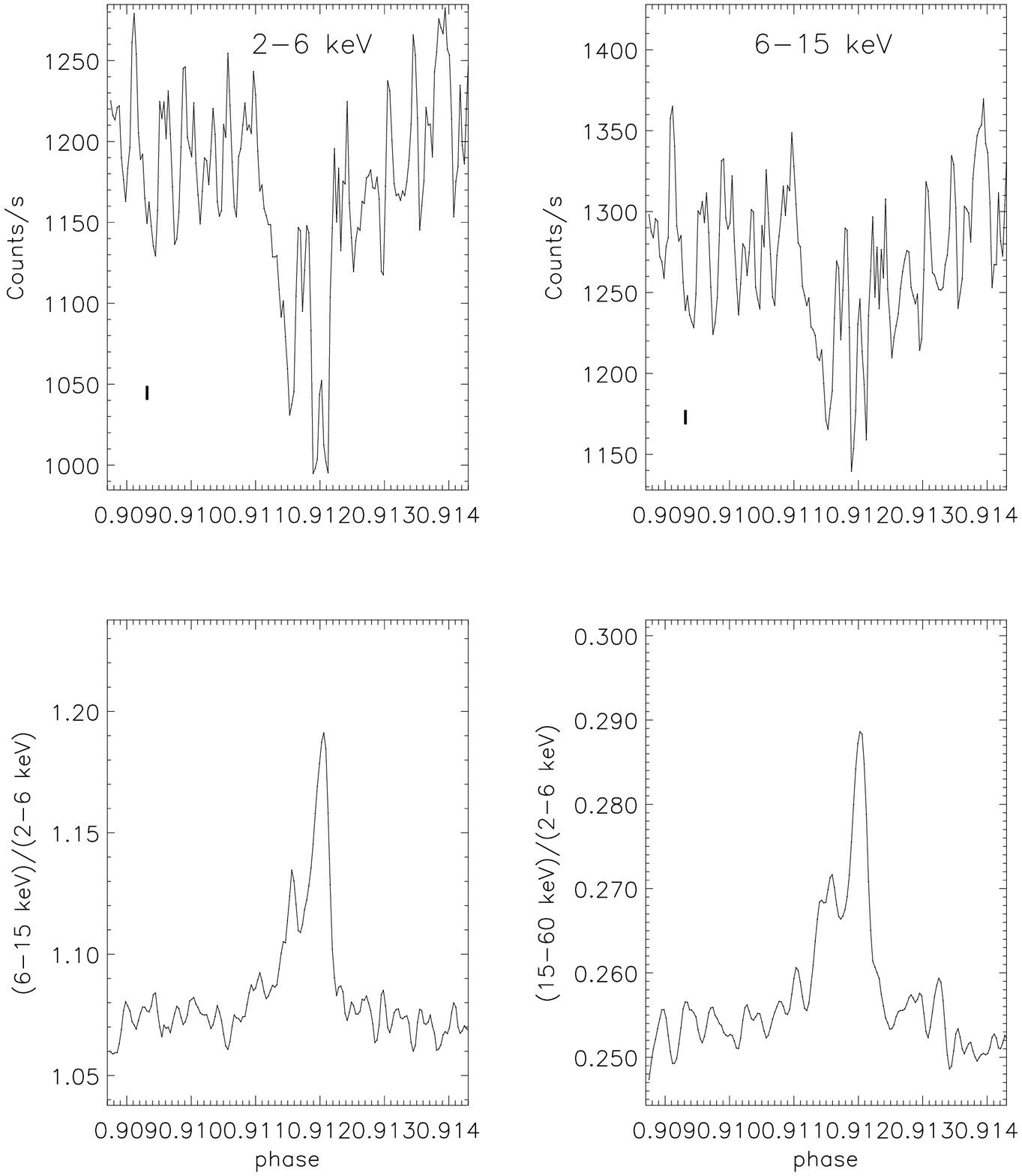,width=1.1\textwidth,angle=0} }
\caption{An example of an asymmetric dip profile. }
\end{figure}
\clearpage
\begin{figure}
\centerline{
\psfig{figure=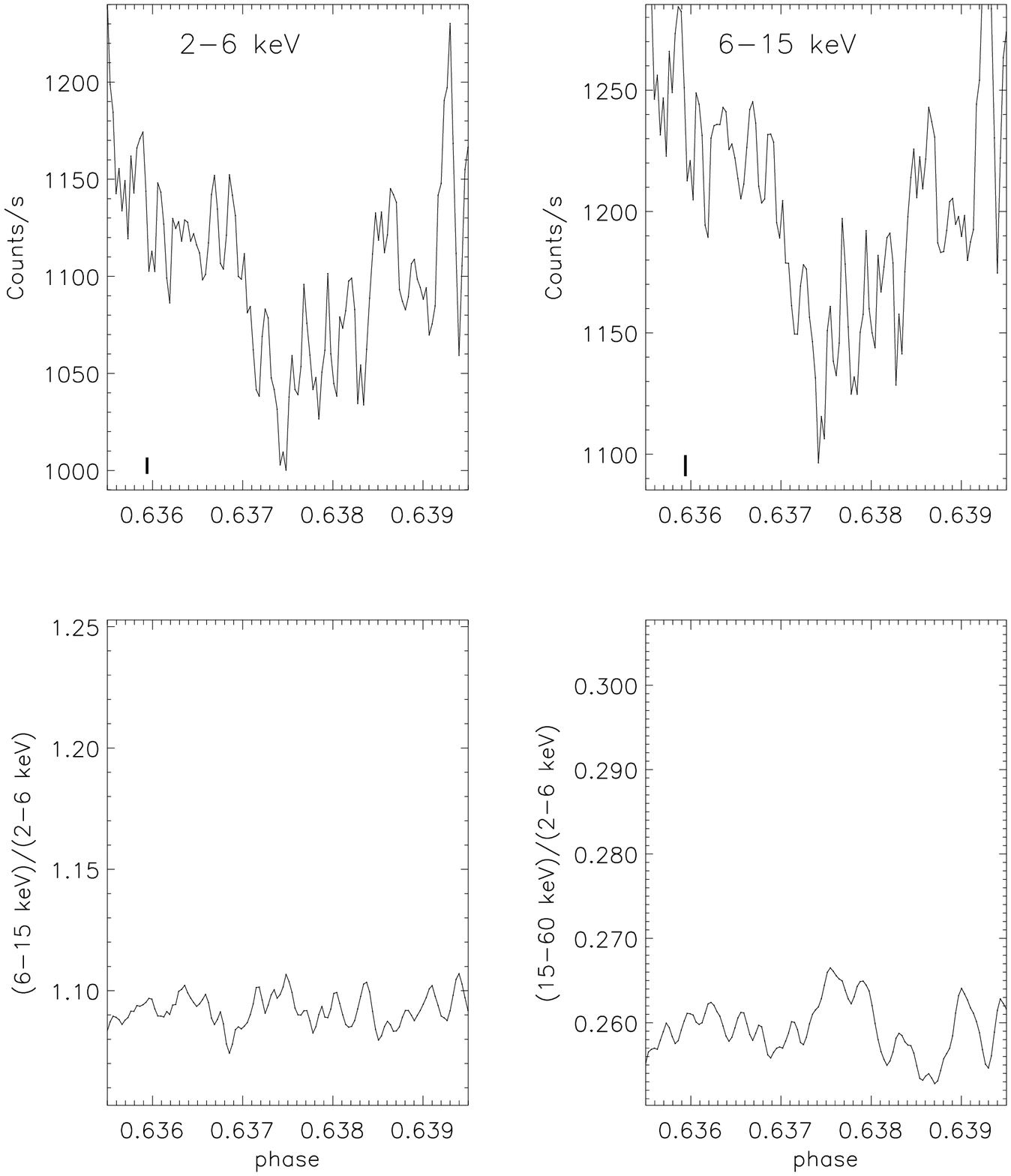,width=1.\textwidth,angle=0} }
\caption{Same as Figure 1, but for a type B dip observed. For comparison,
the vertical scales are adjusted to match those in Figure 1. }
\end{figure}
\clearpage
\begin{figure}
\centerline{
\psfig{figure=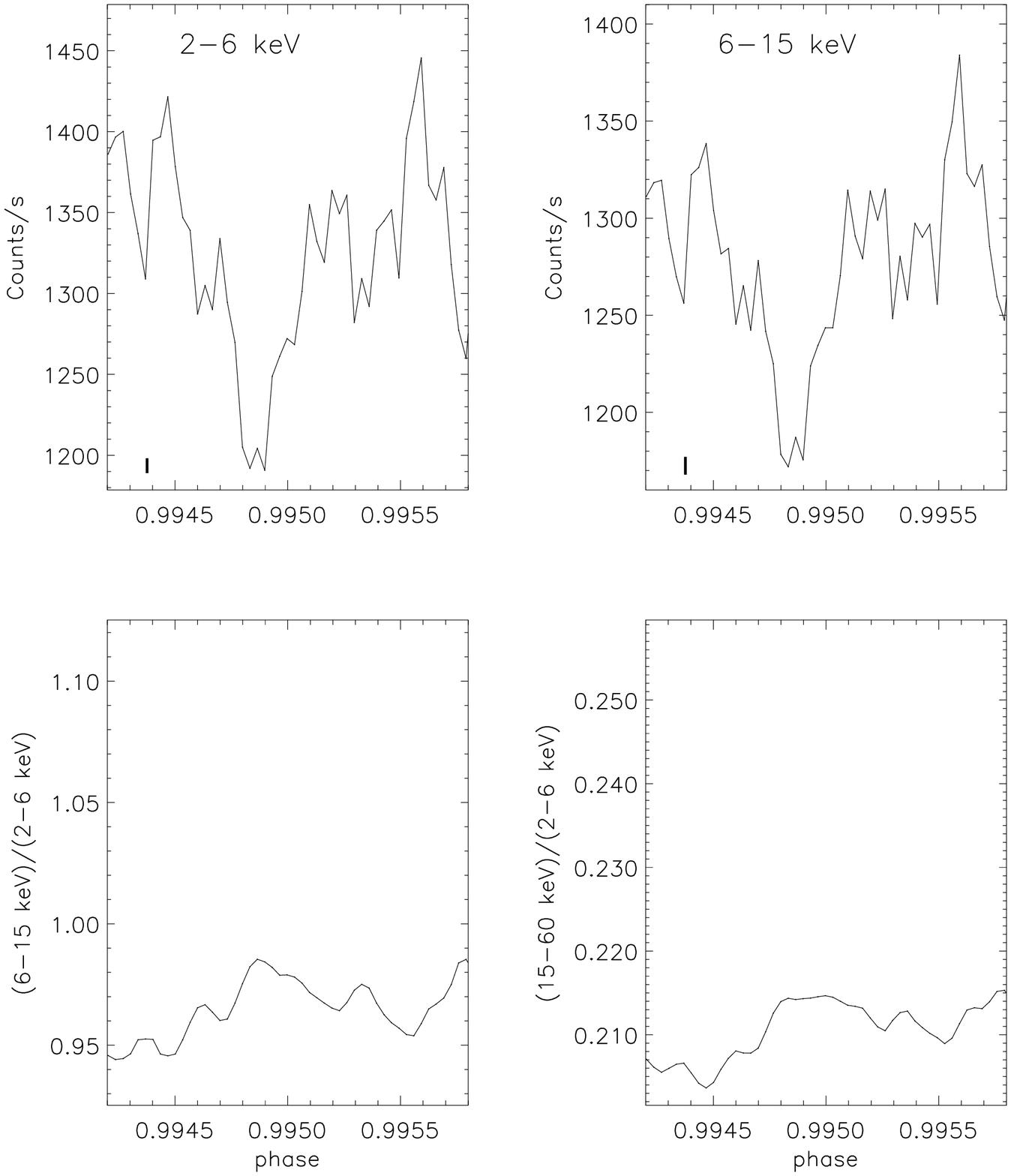,width=1.\textwidth,angle=0} }
\caption{Same as Figure 1, but for a type B dip observed. For comparison,
the vertical scales are adjusted to match those in Figure 1. }
\end{figure}
\clearpage
\begin{figure}
 \centerline{
\psfig{figure=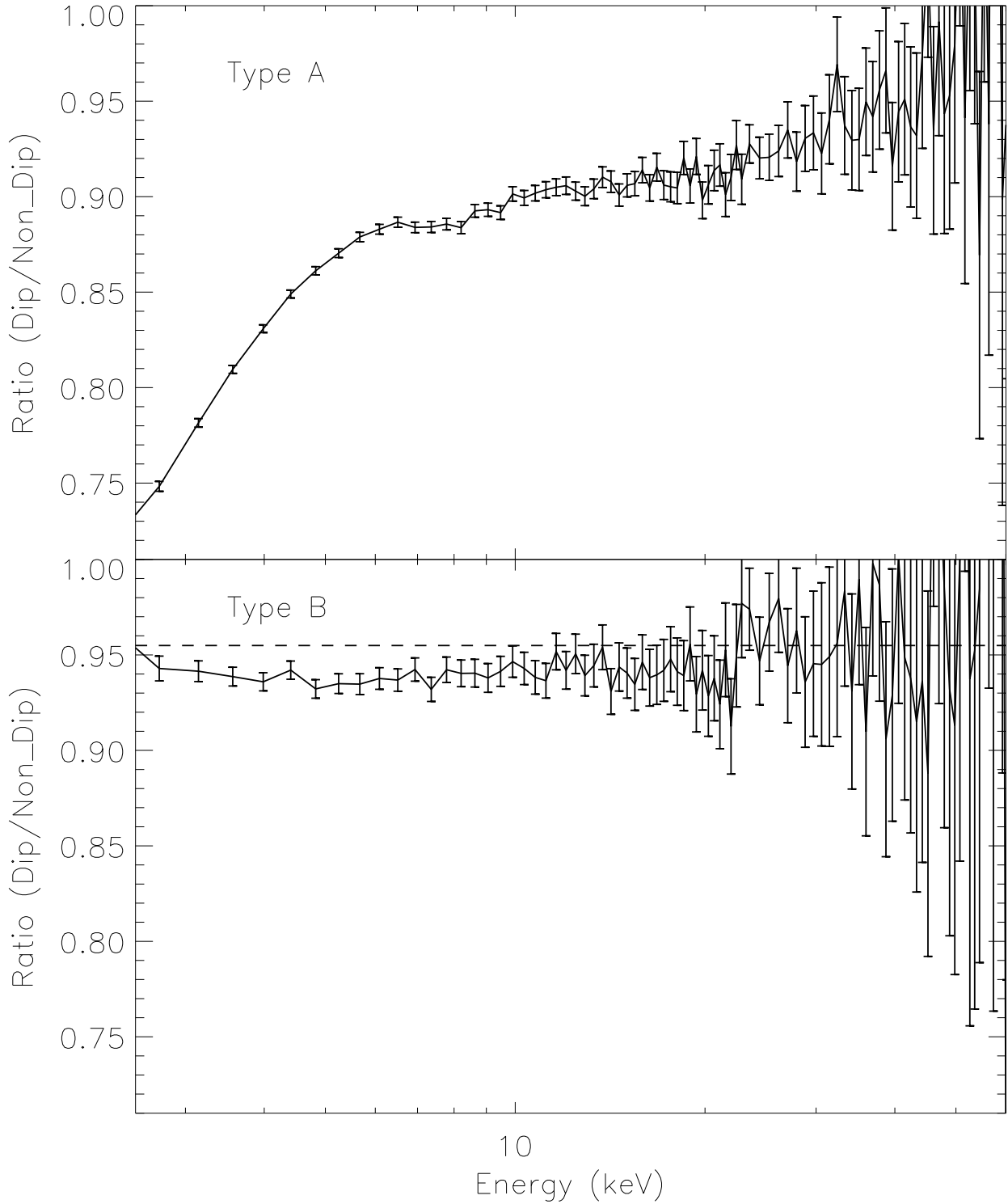,width=1.1\textwidth,angle=0} }
\caption{The ratio of dip spectrum to average non-dip spectrum for a 
type A dip (upper panel) and a type B dip (lower panel), respectively.
The horizontal dashed line in the lower panel represents the average 
value of all points above about 20 keV. }
\end{figure}
\clearpage
\begin{figure}
\centerline{
\psfig{figure=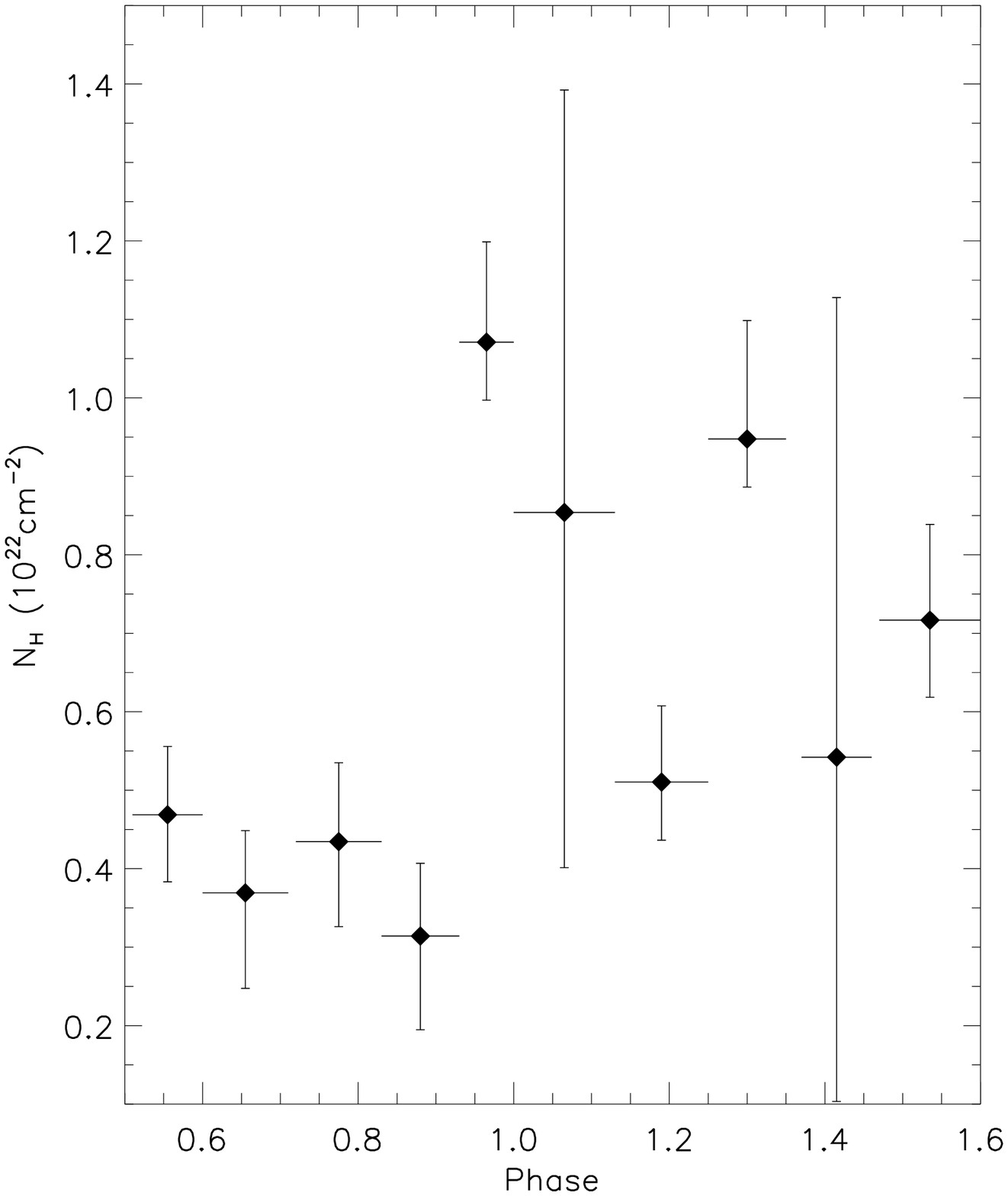,width=1.1\textwidth,angle=0} }
\caption{Observed variation in column density over one orbital cycle. }
\end{figure}


\begin{references}
\reference{}Ba{\l}uci{\'n}ska, M.\ \& Hasinger, G.\ 1991, \aap, 241, 439 
\reference{}Ba{\l}uci{\'n}ska-Church, 
M., Takahashi, T., Ueda, Y., Church, M.\ J., Dotani, T., Mitsuda, K., \& 
Inoue, H.\ 1997, \apjl, 480, L115 
\reference{} Ba{\l}uci{\'n}ska-Church, M., Church, M.\ J., Charles, P.\ A., 
Nagase, F., LaSala, J., \& Barnard, R.\ 2000, \mnras, 311, 861 
\reference{} Blondin, J.\ M., Stevens, I.\ R., \& Kallman, T.\ R.\ 1991,
\apj, 371, 684 
\reference{} Blondin, J.\ M., \& Woo, J. 1995, \apj, 445, 889
\reference{} Bolton,  C. T.  1972, Nature, 235, 271
\reference{} Bolton, C.\ T.\ 1975, \apj, 200, 269 


\reference{} Brocksopp, C., Tarasov, A. E., Lyuty, V. M., \& Roche, P. 1999, 
\aap, 343, 861 
\reference{} Cui, W., Heindl, W.\ A., Rothschild, R.\ E., Zhang, S.\ N., 
Jahoda, K., \& Focke, W.\ 1997, \apjl, 474, L57 
\reference{} Cui, W. 1998, in  ASP. Conf. Ser. 161, High-Energy Processes 
in Accreting Black Holes, ed. J. Poutanen \& R. Svensson, (San Francisco: 
ASP), 97 (astro-ph/9809408)
\reference{} Cui, W., Ebizawa K., Dotani T., and Kubota A., 1998, \apj, 493, 
L75
\reference{} Daniel, J.\ Y.\ 1981, \aap, 94, 121 
\reference{} Ebisawa,  K., Ueda,  Y.,  
Inoue,  H.,  Tanaka,  Y.,   \& White,  N. E.  1996,  \apj,  467,  419
\reference{} Friend, D.~B., \& Castor, J.~I. 1982, ApJ, 261, 293
\reference{} Gies,  D. R. \& Bolton,  C. D.  1986,  \apj,  304,  371
\reference{} Guinan, E.\ F., Dorren, J.\ D., Siah, M.\ J., \& Koch, 
R.\ H.\ 1979, \apj, 229, 296 
\reference{} Herrero, A., Kudritzki, 
R.P., Gabler, R., Vilchez, J. M., \& Gabler, A. 1995, \aap, 297, 556
\reference{} Holt,  S. S.,  Kaluzienski,  
L. J.,  Boldt,  E. A.,   \& Serlemitsos,  P. J.  1979,  \apj,  233,  344
\reference{} Kitamoto, S., 
Miyamoto, S., Tanaka, Y., Ohashi, T., Kondo, Y., Tawara, Y., \& Nakagawa, 
M.\ 1984, \pasj, 36, 731 
\reference{} Liang,  E. P.  \& Nolan,  P. L., 1984, Space Sci. Rev, 38, 353
\reference{} Liang,  E. P. 1998, Physics Reports, 302, 67
\reference{} Long, K. S., Chanan, G. A., \& Novick, R. 1980, \apj, 238, 710
\reference{} Oda, M. 1977, Space Sci. Rev., 20, 757
\reference{} Pravdo,  S. H.,  White,  N. 
E.,  Becker,  R. H.,  Kondo,  Y.,  Boldt,  E. A.,  Holt,  S. S., 
 Serlemitsos,  P. 
J.,   \& McCluskey,  G. E.  1980,  \apjl,  237,  L71
\reference{} Priedhorsky, W. C.,  Brandt,  S.,  Lund,  N.  1995,  \aap,  
300,  415
\reference{} Remillard, R.\ 
A.\ \& Canizares, C.\ R.\ 1984, \apj, 278, 761 
\reference{} Schulz, N. S., Cui, W., Canizares, C. R., Marshall, H. L.,
Lee, J. C., Miller, J. M., \& Lewin, W. H. G. 2001, ApJ, submitted
\reference{} 
Tanaka,~Y., \& Lewin,~W.~H.~G. 1995, in X--ray Binaries, ed. 
W. H. G. Lewin, J. van Paradijs, \& E. P. J. van den Heuvel
(Cambridge U. Press, Cambridge), 126
\reference{} Walborn, N.R., 1973, AJ, 78, 10, 1067
\reference{} Webster, B. L., \& Murdin, P. 1972, Nature, 235, 37
\reference{} Wen, L., Cui, W., Levine, A.\ M., \& Bradt, H.\ V.\ 1999, 
\apj, 525, 968 
\reference{} Wen, L., Cui, W., \& Bradt, ~H.~V. 2001, \apj, 546, 105
\end{references}
\end{document}